\documentclass[twocolumn,superscriptaddress,prb,aps,preprintnumbers,longbibliography]{revtex4-2}

\usepackage{amsmath,amssymb,bm,graphicx,color,gensymb,bbold,appendix,xspace}
\usepackage[colorlinks=true, citecolor={blue!80!black}, urlcolor={blue!50!black}, linkcolor = {blue!80!black}]{hyperref}
\usepackage[utf8x]{inputenc}

\usepackage{xcolor}

\newcommand{\mub}{\mu_\mathrm{B}}

\newcommand{\SCB}{SrCu$_2$(BO$_3$)$_2$\xspace}

\newcommand{\CsCuCl}{Cs$_2$CuCl$_4$\xspace}

\newcommand{\CsCoCl}{Cs$_2$CoCl$_4$\xspace}
\newcommand{\CsCoBr}{Cs$_2$CoBr$_4$\xspace}

\newcommand{\ii}{\mathrm{i}}

\newcommand{\be}{\begin{equation} }
	\newcommand{\ee}{\end{equation} }
\newcommand{\bea}{\begin{eqnarray} }
	\newcommand{\eea}{\end{eqnarray} }
\renewcommand{\mub}{\mu_\mathrm{B}}

\newcommand{\ham}[1]{\hat{\mathcal{#1}}}
\newcommand{\mb}[1]{\mathbf{#1}}

\begin{document}

\title{Magnetic field-induced phases and spin Hamiltonian in \CsCoBr}

\author{L.~Facheris}
\address{Laboratory for Solid State Physics, ETH Z\"{u}rich, 8093 Z\"{u}rich, Switzerland}
\author{S. D.~Nabi}
\address{Laboratory for Solid State Physics, ETH Z\"{u}rich, 8093 Z\"{u}rich, Switzerland}
\author{K.~Yu.~Povarov}
\address{Laboratory for Solid State Physics, ETH Z\"{u}rich, 8093 Z\"{u}rich, Switzerland}
\address{Present address: Dresden High Magnetic Field Laboratory (HLD-EMFL) and W\"urzburg-Dresden Cluster of Excellence ct.qmat, Helmholtz-Zentrum Dresden-Rossendorf, 01328 Dresden, Germany}
\author{Z. Yan}
\address{Laboratory for Solid State Physics, ETH Z\"{u}rich, 8093 Z\"{u}rich, Switzerland}

\author{A.~Glezer~Moshe}
\author{U.~Nagel}
\author{T.~R{\~o}{\~o}m}
\address{National Institute of Chemical Physics and Biophysics, Akadeemia tee 23, 12618 Tallinn, Estonia}

\author{A.~Podlesnyak}
\address{Neutron Scattering Division, Oak Ridge National Laboratory, Oak Ridge, Tennessee 37831, USA}

\author{E.~Ressouche}
\author{K.~Beauvois}
\address{Université Grenoble Alpes, CEA, IRIG, MEM, MDN, 38000 Grenoble, France}

\author{J. R. Stewart}
\author{P.~Manuel}
\author{D.~Khalyavin}
\author{F.~Orlandi}
\address{ISIS Neutron and Muon Source, Rutherford Appleton Laboratory, Didcot, OX11 0QX, United Kingdom}

\author{A.~Zheludev}
\email{zhelud@ethz.ch; http://www.neutron.ethz.ch/}
\address{Laboratory for Solid State Physics, ETH Z\"{u}rich, 8093 Z\"{u}rich, Switzerland}

\begin{abstract}
Magnetic structures and spin excitations are studied across the phase diagram of the geometrically frustrated $S=3/2$ quantum antiferromagnet \CsCoBr in magnetic fields applied along the magnetic easy axis, using neutron diffraction, inelastic neutron scattering and THz absorption spectroscopy. The data are analyzed, where appropriate, using extended $SU(4)$ linear spin wave theory. A minimal magnetic Hamiltonian is proposed based on measurements in the high field paramagnetic state. It deviates considerably from the previously considered models. Additional dilatometry experiments highlight the importance of magnetoelastic coupling in this system.  
\end{abstract}
\date{\today}
\maketitle

\section{Introduction}
The title compound is a member of a well-known family of frustrated triangular antiferromagnets that includes such thoroughly-studied systems as \CsCuCl \cite{Coldea2001,Coldea2002,Coldea2003,Tokiwa2006}, Cs$_2$CuBr$_4$ \cite{Ono2003,Fortune2009} and \CsCoCl \cite{Kenzelmann2002,Breunig2013,Breunig2015}. Nevertheless, recent magnetic, thermodynamic and neutron scattering studies of \CsCoBr  have demonstrated its uniqueness \cite{PovarovFacheris2020,Facheris2022,Facheris2023}. The excitation spectrum is a complex hierarchy of bound states of fractional excitations, reminiscent of Zeeman ladders in Ising spin chains \cite{McCoy1978,Shiba1980,Rutkevich2008,Coldea2010}.  Indeed, the magnetic lattice in \CsCoBr is a distorted triangular one, with interactions along one particular crystallographic direction somewhat stronger than the transverse coupling.  However, the situation in \CsCoBr is far more complex than in the one-dimensional (1D) Ising model. The material features dominant easy-plane, rather than Ising anisotropy. The observed bound states are not confined to single ``chains'', and also propagate in two dimensions \cite{Facheris2023}. In applied magnetic fields \CsCoBr demonstrates a complex phase diagram with as many as five distinct ordered phases \cite{PovarovFacheris2020,Facheris2022}. Apart from the Néel ``stripe'' ground state, those include an incommensurate spin density wave (SDW)  phase and a commensurate fractional magnetization plateau.

Despite these rather spectacular experimental findings, the physics of \CsCoBr remains poorly understood. To date, the spin Hamiltonian could only be guessed from bulk measurements and a crude spin wave theory (SWT) analysis of neutron spectra collected in zero applied field \cite{PovarovFacheris2020,Facheris2022}. The obvious limitation of this approach is that this spectrum is Zeeman-ladder like and actually has little resemblance of anything that semiclassical SWT predicts. Another limitation is the complete frustration of inter-chain interactions in the stripe phase. This suppresses the transverse dispersion of magnetic excitations and makes it insensitive to any details of inter-chain coupling. The third is that the quantum nature of spins involved and reduced dimensionality most likely result in a substantial re-normalization of exchange constants: even if the spectrum were adequately described by SWT, the values of the exchange constants obtained in any fit to the data would be significantly different from the actual constants in the Hamiltonian. The final limitation of the previous analysis is the reliance on a pseudospin-$1/2$ projection of the actual $S=3/2$ Hamiltonian appropriate for describing the Co$^{2+}$ ions in this material \cite{Breunig2013,Facheris2022}. The approximation would be excellent if the single-ion anisotropy were orders of magnitude stronger than exchange interactions and Zeeman energies of applied fields, but that is not quite the case for \CsCoBr. All in all, one has to admit that the spin Hamiltonian remains unknown.

The second gap in our understanding is of a technical nature. Due to the geometry of split-coil magnets used in neutron experiments, it is rather difficult to simultaneously apply a large magnetic field along the chain axis and measure at large momentum transfers along that same direction. Unfortunately, it is precisely this type of experiment that is needed to accurately determine the magnetic structures and to understand the spin dynamics in applied fields. As a result, the nature of the two high-field phases has not been clarified to date.

The present work aims to address these issues. We report comprehensive magnetic diffraction and spectroscopy experiments on \CsCoBr in applied magnetic fields. First, we determine the magnetic structures of the two remaining high-field ordered phases, thereby completing the exploration of the $H-T$ phase diagram. In this context we also report dilatometric measurements that reveal the importance of magneto-elastic effects across the phase boundaries. Next, we apply generalized $SU(4)$ SWT to analyze the neutron and THz spectra collected in the high-field paramagnetic state. Since SWT is known to work well in that regime, it provides reliable estimates of the actual exchange constants. The results lead us to a substantial revision of the ``minimal model'' for \CsCoBr. Finally, we use the exchange parameters measured in high field to make sense of the spectra collected in several low-field phases, including the plateau state. 

\section{Structural considerations and minimal model}
Before describing the experiments performed in this work we summarize the basic facts about the crystal structure of \CsCoBr and consider the single-ion magnetic anisotropy and possible exchange interactions.

 As discussed in some detail in Refs.~\cite{PovarovFacheris2020,Facheris2022}, the material is orthorhombic, space group  $Pnma$, with lattice parameters $a=10.137(1)$~\AA\, $b=7.593(3)$~\AA\ and  $c=13.281(1)$~\AA\ at $\sim 100$~mK. There are four equivalent magnetic Co$^{2+}$ ions in the unit cell, each encased in a Br$_4$ tetrahedron. The point symmetry of the Co site allows for general single-ion anisotropy and gyromagnetic tensors with the restriction that one principal axis of the corresponding ellipsoid must coincide with the $\mb{b}$ direction. A previous analysis of high-temperature susceptibility data suggested that the single-ion anisotropy term in the spin Hamiltonian is of predominantly the XY-type. The ``hard axis'' $\mb{z}$ is perpendicular to the  crystallographic $\mb{b}$ axis (denoted as $\mb{y}$), and forma an angle $\beta \approx 44^\circ$ with the $\mb{a}$ direction. For the  $S=3/2$ ion Co$^{2+}$ the most general symmetry-compatible single-ion Hamiltonian is
 \be
 \ham{H}_\text{single-ion}=-D{\hat{S}_x}^2 -(D+d){\hat{S}_y}^2,\label{eq:anis0}
 \ee 
where $d$ is a small Ising-like contribution, always along the chains, i.e., the crystallographic $\mb{b}$-axis.

\begin{figure}
	\includegraphics[width=\columnwidth]{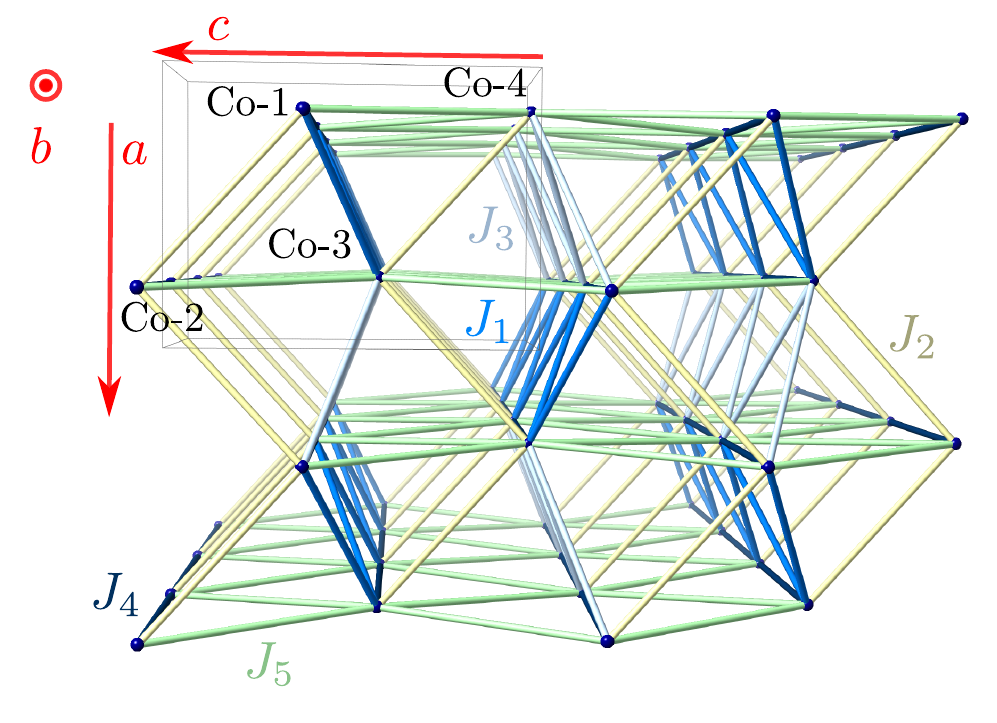}
	\caption{The five nearest-neighbor magnetic interactions in \CsCoBr. The nodes of the lattice are Co$^{2+}$ ions. The boundaries of a unit cell are shown in black lines in the top-left corner. }\label{fig:coupling}
\end{figure}

The principal antiferromagnetic exchange constant referred to as $J$ in Refs.~\cite{PovarovFacheris2020,Facheris2022,Facheris2023} connects nearest neighbors along the $\mb{b}$-direction, forming the above-mentioned chains. This interaction is the only one corresponding to an obvious Co-Br-Br-Co supercharge pathway with a short (3.71~\AA) Br-Br distance. Yet, at $7.593$~\AA~$=b$ it is only the 4-th shortest Co-Co bond, and will thus be denoted below as $J_4$. Being a principal crystallographic translation, this bond connects two CoBr$_4$ tetrahedra  with identical orientations. The anisotropy axes of the corresponding interacting ions coincide.

As in the much-studied \CsCuCl \cite{Coldea2001}, the distorted triangular lattice is completed by a bond that  is almost parallel to the $(b,c)$ plane and connects magnetic sites in adjacent chains. We previously denoted it as $J'$ \cite{PovarovFacheris2020,Facheris2022,Facheris2023}.  It happens to be the 5th-shortest Co-Co distance of 7.65~\AA\ and will be henceforth denoted as $J_5$. The coordination tetrahedra of the  Co$^{2+}$ ions that it connects are related by a glide symmetry. The normal vectors to the corresponding anisotropy planes form angles of $\pm \beta$ with the $\mb{a}$ axis, which makes them almost orthogonal.

As will be shown below, these two coupling constants are not sufficient to fully describe the dispersion of magnetic excitations in \CsCoBr. We therefore also consider interactions across the three shortest Co-Co distances, as shown in Fig.~\ref{fig:coupling}. The nearest-neighbor bond associated with $J_1$ is only $6.50$~\AA\ long. I connects adjacent chains (and with them, the adjacent triangular planes) along the $\mb{a}$ direction, in a zigzag manner. Together with $J_4$ it forms zigzag ladders running along the $\mb{b}$ axis. Just like $J_5$, this bond is frustrated by the AF coupling within each chain. The sites that it connects are related by inversion, and therefore have identical single-ion anisotropy and gyromagnetic tensors.

A very similar connection between triangular planes is provided by $J_3$ that spans over  6.86~\AA.  Together with $J_1$ and $J_4$, $J_3$ completes a somewhat buckled triangular lattice in the $(a,b)$ plane. Unlike in the previously discussed $(b,c)$-triangular lattice, the inter-chains bonds $J_3$ and $J_1$ are {\em not equivalent}. It will be later argued that one of them must be negligible compared to the other.

The last potential Co-Co exchange constant that we shall consider here is $J_2$. It spans over 6.79~\AA\  and also connects adjacent $(b,c)$-triangular planes. Unlike all other inter-chain bonds discussed above,  it is {\em not frustrated} by antiferromagnetic in-chain $J_4$-interactions.  However, like $J_5$, it connects ions with almost orthogonal anisotropy planes.

As mentioned, there are good microscopic reasons for $J_4$ to be strong, but the hierarchy of the remaining exchange constants is hard to guess based on just the crystal structure. We note, however, that the next (6th) coupling would have to span over a distance of more than 10~\AA. Our ``minimal model'' for \CsCoBr will therefore be restricted to just $J_1$--$J_5$.

By symmetry considerations alone, all the above-mentioned interactions may include anisotropic exchanges and also some off-diagonal Dzyaloshinskii-Moriya terms. Taking all that into account would result in 45 independent exchange parameters and clearly an impractical model. Instead we shall assume that all interactions are of {\em Heisenberg} type, and that the single-ion term (\ref{eq:anis0}) is the only source of magnetic anisotropy in \CsCoBr. While the assumption is clearly arbitrary, below we show that this minimal model is able to adequately describe many features of the spin excitation spectrum in a wide range of applied fields.

\section{Experimental procedures}

Magnetic propagation vectors in the high-field phases were determined with the WISH time-of-flight diffractometer at Rutherford Appleton Laboratory (RAL, UK) \cite{WISH_paper}. The sample environment consisted of a $^3$He-$^4$He dilution refrigerator
in combination with a $10$~T vertical superconducting cryomagnet. A $221.4$~mg single crystal was installed with the $(a,c)$ plane horizontal and the field applied along the $\mb{b}$ axis. Thanks to the large detector bank, the propagation vector of the “D” and “E” phases could be identified \cite{WISH_data}.

Following that, extensive nuclear scattering and magnetic diffraction data sets were collected on the CEA-CRG D23 lifting-counter diffractometer at Institut Laue-Langevin (ILL, France). The $36.6$~mg single crystal was mounted on the cold finger of a $^3$He-$^4$He dilution refrigerator in a $6$~T superconducting cryomagnet. The field was applied vertically, along the crystallographic $\mb{b}$ axis. Sample mosaic was approximately $0.5^\circ$~FWHM. All the measurements were performed at the base temperature of the dilution insert $< 100$~mK. Neutrons with $\lambda=1.28$~\AA~(copper monochromator) and $2.39$~\AA~(pyrolytic graphite monochromator) were used for nuclear and magnetic studies, respectively. 

Inelastic neutron scattering measurements in applied fields were conducted on the CNCS time-of-flight spectrometer at Oak Ridge National laboratory (ORNL, USA) \cite{CNCS_paper}, using $3.32$~meV incident energy neutrons. The $825$~mg single crystal sample was aligned with the $\mb{b}$ axis vertical. The sample environment consisted of a $^3$He-$^4$He dilution refrigerator and an $8$~T vertical field cryomagnet ($\mb{H}\|\mb{b}$ at all times). The spectrum was measured by rotating the sample over a $190^\circ$ angular range around the $\mathbf{b}$ axis in steps of $1^\circ$. Typical counting times were $\sim 5$ minutes at each sample orientation. Energy resolution was $\approx 0.10$~meV at the elastic position. All the measurements were performed at the base temperature of the dilution insert $\sim 100$~mK. The data were reduced using MANTID \cite{MANTID_paper} and HORACE \cite{HORACE_paper} software.

Below we also show some previously unpublished data from a zero-field inelastic neutron experiment carried out at the LET spectrometer at RAL~\cite{Facheris2023,LET_data}. In this measurement we used the same $1.16$~g single crystal as Refs.~\cite{Facheris2022,Facheris2023}. The sample was mounted with the $\mb{a}$ axis vertical. All measurements were performed at the base temperature of a $^3$He-$^4$He dilution refrigerator. The data were collected as the sample was rotated over a $180^\circ$ angular range around the vertical axis, in steps of $1^\circ$, typically counting $\sim 10$ minutes at each orientation. The data were reduced with MANTID \cite{MANTID_paper} and HORACE \cite{HORACE_paper}.

The THz absorption experiment was performed with a Martin-Puplett-type interferometer and a $^3$He-$^4$He dilution refrigerator using a $^3$He-cooled Si bolometer at $300$~mK. The sample was a circular plate $1.1$~mm thick in the $\mb{b}$ direction and $4$~mm in diameter. THz radiation propagating along the crystal $\mb{b}$ axis was unpolarized and the apodized instrumental resolution was $0.025$~meV. Due to the heating caused by the THz radiation, the lowest stable attainable temperature was around $200$~mK.

Studies of magnetostriction were performed by measuring the capacitance between two electrodes deposited as thin layers of conducting Silver paste on the surface of a plate-shaped sample. Thin coaxial cables were attached to the electrodes and carried the signal to an Andeen-Hagerling A2550A ultra-precision capacitance bridge operating with a $1$~kHz AC probing field and an excitation voltage of $15$~V. The measurements were performed with a $^3$He-$^4$He dilution refrigerator insert for a commercial Quantum Design Physical Property Measurement System (PPMS) equipped with a $9$~T superconducting magnet. In the experiment, the capacitance was tracked versus increasing field for a set of fixed temperatures. The field was swept at $10$~Oe/s and the raw data were averaged over $200$~Oe wide bins.  In all the cases, the probing electric field was $\mathbf{E}\parallel\mathbf{a}$, while the sample was installed with $\mathbf{H}\parallel\mathbf{b}$. Only the real part of the complex capacitance is discussed, as it showed a modulated response versus field, in contrast to the imaginary part which was in all the cases flat.

\section{Results and data analysis}
We now report and briefly discuss the results of different experiments in applied magnetic fields. In all cases, the field was directed along the crystallographic $\mb{b}$ axis.

\subsection{Low-temperature crystal structure}

The crystal structure in zero field and a temperature of $\sim 0.1$~K was previously determined in a neutron diffraction experiment at the D23 lifting-counter diffractometer at ILL. Details of the measurements and structural refinement are given in the Supplement of Ref.~\cite{Facheris2022}. For the sake of completeness,  here we provide the resulting atomic coordinates, which were not reported to date. These are listed in Table \ref{LowTStruct}.

\begin{table*}
	\caption{\label{LowTStruct} Crystal structure parameters of Cs$_2$CoBr$_4$ at $100$~mK determined through neutron diffraction, as described in the Supplement of Ref.~\cite{Facheris2022}.}
	\begin{ruledtabular}
		\begin{tabular}{ccccc}
			Atom & $x$ & $y$ & $z$ & $B_{\text{iso}}$ \\
			\hline
			Cs & 0.5240(6) & 0.75 & 0.3339(5) & 0.59(12) \\
			Cs & 0.8604(8) & 0.75 & 0.6002(6) & 0.53(15) \\
			Co & 0.2616(12) & 0.75 & 0.5801(10) & 0.73(27) \\
			Br & 0.4970(6) & 0.75 & 0.6055(5) & 0.73(11) \\
			Br & 0.1901(9) & 0.75 & 0.4050(4) & 0.74(11) \\
			Br & 0.1694(4) & 0.4946(29) & 0.65279(16) & 0.88(5) \\
		\end{tabular}
	\end{ruledtabular}
\end{table*}

\subsection{Magnetic neutron diffraction}
\subsubsection{Succession of magnetic propagation vectors}

\begin{figure}
	\includegraphics[width=\columnwidth]{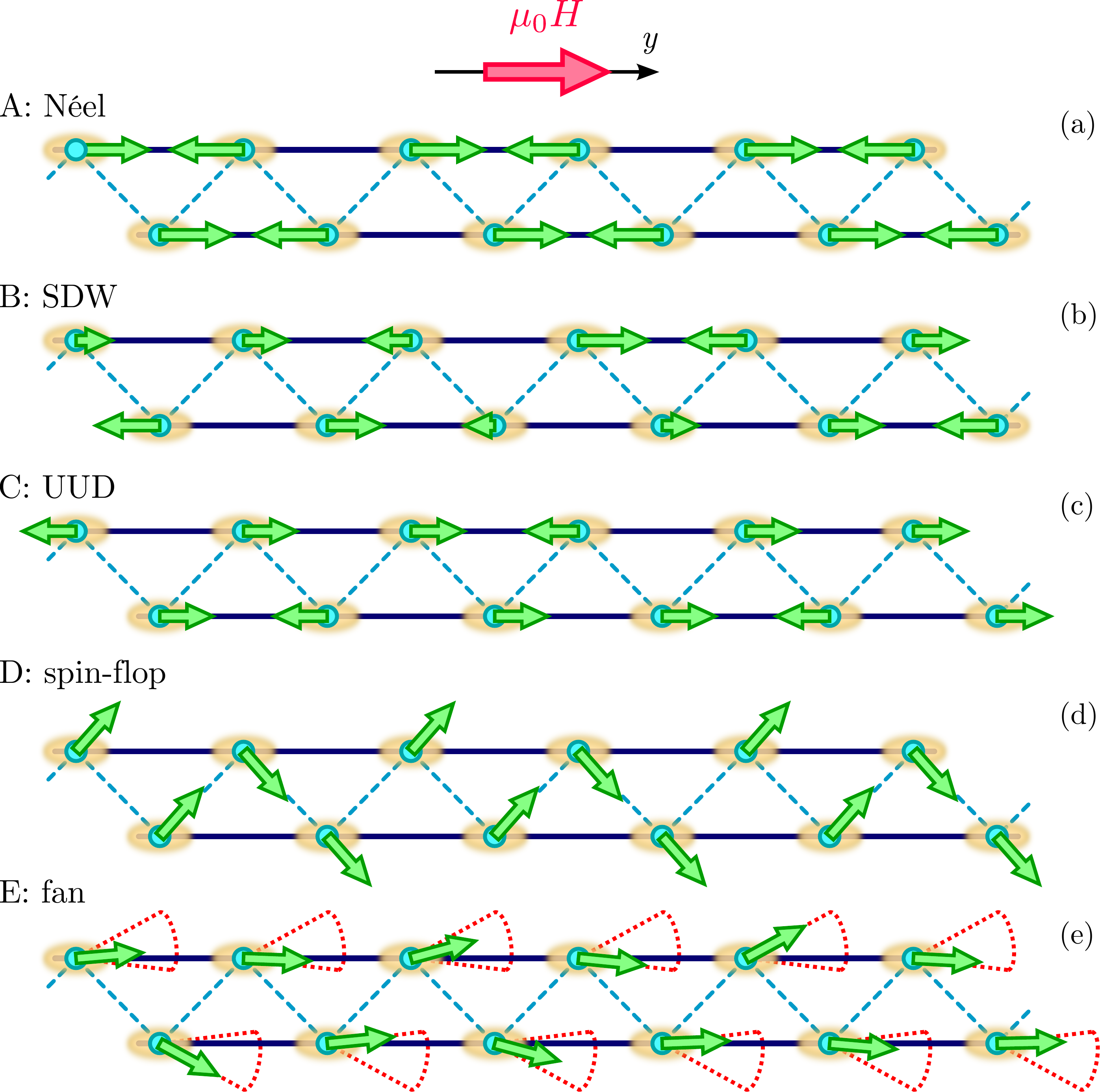}
	\caption{\label{fig:structure} A schematic representation of magnetic structures that occur in \CsCoBr in magnetic fields applied along the $\mb{b}$ axis. Only the Co-1 and Co-3 sites that form $J_4-J_1$ zigzag ladders are shown. For these, the individual magnetic moments are in all cases confined to a plane which coincides with the local anisotropy easy plane common to Co-1 and Co-3.  The red dashed lines show the range of transverse spin oscillations in this incommensurately modulated structure}. The ellipses represent additional weak in-plane anisotropy (parameter $d$ in Eq.~(\ref{eq:anis0})).
\end{figure}

As discussed in detail in Ref.~\cite{PovarovFacheris2020}, in \CsCoBr applying a magnetic field along the crystallographic $\mb{b}$ axis causes a cascade of transitions between magnetic phases A-F at $H_\text{AB}= 1.5$~T, $H_\text{BC}= 2.7$~T, $H_\text{CD}= 4.0$~T, $H_\text{DE}= 4.5$~T, and $H_\text{sat}=5.2$~T. The corresponding phase boundaries in the $(H,T)$ plane are reproduced in symbols in Fig.~\ref{fig:dilation} below. All phase transitions except the one at $H_\text{sat}$ are discontinuous. The magnetic structures of phases A, B and C have been previously solved from neutron diffraction \cite{Facheris2022}: Phase A is the Néel or ``stripe'' phase with propagation vector $(0,1/2,1/2)$. Phase B is an incommensurate longitudinal spin density wave (SDW). The propagation vector is $(0,\xi,0)$ and is strongly field-dependent, with $\xi$ proportional to the $S=1/2$ pseudospin-magnetization. The pseudospin-plateau Phase C is a lock-in version of Phase B, with propagation vector $(0,1/3,0)$. It can be seen as a classical ``up-up-down'' (UUD) structure. All this is illustrated in Fig.~\ref{fig:structure}(A-C), where only the Co-1 and Co-3 sites belonging to a single $J_4-J_1$ zigzag ladder are shown. Beyond $H_\text{sat}$  is the paramagnetic ``pseudospin-saturated'' state.

\begin{figure}
	\includegraphics[width=\columnwidth]{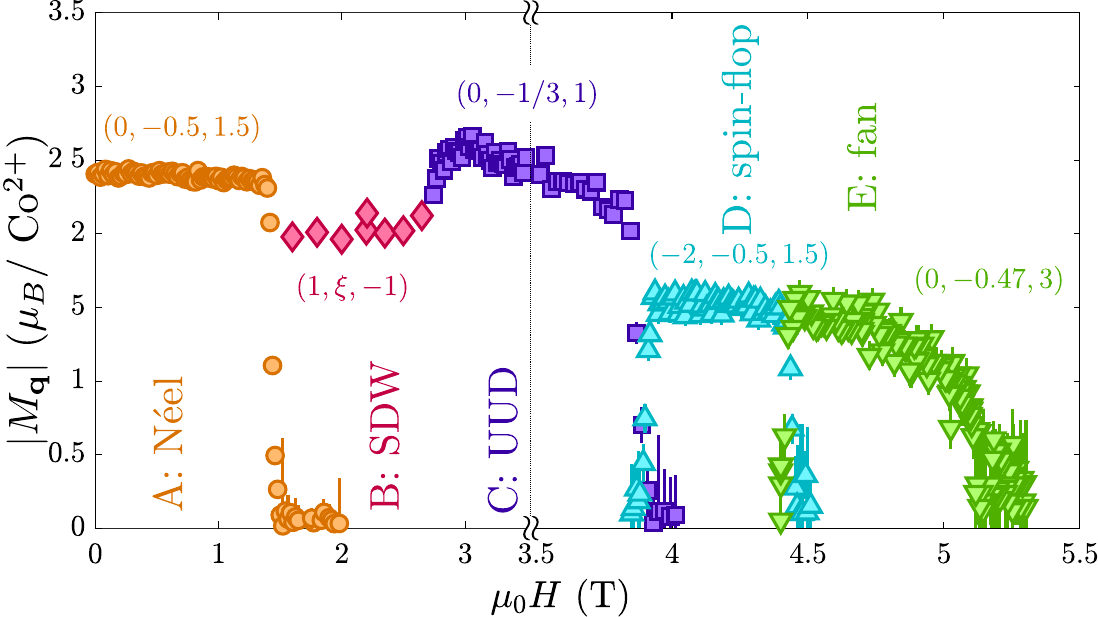}
	\caption{\label{Fig_BraggvsH} Magnetic order parameter (amplitude of the corresponding Fourier component of magnetization) versus magnetic field applied along the $\mb{b}$ axis, as deduced from the intensities of the indicated Bragg reflections measured in \CsCoBr at $T<100$~mK. The data for Phases A-C are from Ref.~\cite{Facheris2022}.}
\end{figure}

Measurements on the WISH instrument at RAL revealed the propagation vectors in the two remaining phases: $(0,1/2,1/2)$ for Phase D and $(0,0.47,0)$ in Phase E. Subsequent measurements on the D23 diffractometer have shown that the propagation vector in Phase E is {\em field-independent}, unlike Phase B. In Fig.~\ref{Fig_BraggvsH} we show the evolution of the magnetic order parameters in each phase. For Phases A-C the data are from \cite{Facheris2022}, and for Phases D and E from the analysis described below.

\subsubsection{Description of the magnetic structures}
There are four Co sites in each unit cell. Their cell coordinates are Co-1: $(0.262(1), 0.75, 0.580(1))$, Co-2: $ (0.762(1), 0.75, 0.920(1))$, Co-3: $ (0.738(1), 0.25, 0.420(1))$ and Co-4: $ (0.238(1), 0.25, 0.080(1))$, following the convention in \cite{Facheris2022}. In the propagation vector formalism, the part of the magnetization that is modulated with periodicity $\mb{q}$ is described in terms of complex amplitude vectors $\mb{A}^{(\mb{q})}_{d}$, $d=1..4$, for each ion in the 1st unit cell, respectively. The contribution to the magnetic moment $\mb{m}^{(\mb{q})}_{i,d}$ of the $d$-th ion in the $i$-th unit cell with origin at $\mb{R}_i$ can be written as:
\be
\mb{m}^{(\mb{q})}_{i,d}=\frac{1}{2}\mb{A}^{(\mb{q})}_{d}\,\exp\left[\ii \mb{q}\mb{R}_i\right]+\text{c.c.}
\ee
In this way the modulated part of magnetization is parameterized by $3\times4$ complex vector components. These parameters are restricted to comply with crystallographic symmetry. Specifically, they are chosen as linear combinations of basis vectors of irreducible representations of the ``little group''~\cite{SARAh_paper}. The latter is the subgroup of the crystallographic space group that leaves the propagation vector intact. The coefficients $C_1$, $C_2,$,.. of these linear combinations are fit to reproduce the measured Bragg intensities.

In the presence of an external magnetic field there is an {\em additional} contribution to the actual magnetic moment $\mb{m}_{i,d}$ of each ion, namely the one corresponding to the propagation vector $(0,0,0)$: $\mb{m}_{i,d}=\mb{m}^{(\mb{q})}_{i,d}+\mb{m}^{(0)}_{i,d}$. $\mb{m}^{(0)}_{i,d}$ is expanded  in  the same manner as $\mb{m}^{(\mb{q})}_{i,d}$, which introduces additional $3\times4$ real parameters, the components of $\mb{A}^{(0)}_{d}$. These can {\em not} be reliably measured in an unpolarized neutron diffraction experiment, because the weak magnetic Bragg reflections with propagation vector $(0,0,0)$ overlap with the much stronger nuclear Bragg peaks. Only the $(0,0,0)$ ``Bragg peak'' is accessible experimentally: its ``intensity'' is the square of the total magnetization of the sample. Therefore, reconstructing the actual magnetic structure $\mb{m}_{i,d}$ from a diffraction experiment is only possible if additional assumptions  regarding the vectors $\mb{A}^{(0)}_{d}$ are made, as described below.

\subsubsection{Spin-flop phase}
The spin structure at 4.25~T in Phase D, where the propagation vector is $(0,1/2,1/2)$, was determined from the analysis of 70 magnetic Bragg intensities. These were collected  in rocking curves with a 0.06$^\circ$ step and counting $15$~s/point. The symmetry-based group theory analysis was performed using the SARA$h$ software~\cite{SARAh_paper}. It restricts the possible structure to the same two irreducible representations as in zero field  \cite{Facheris2022}. Of these only $\Gamma_2$ is consistent with the measured pattern of intensities, as is also the case in zero field. Its basis vectors are tabulated in Table \ref{irrepD}. 

\begin{table*}
	\begin{ruledtabular}
		\begin{tabular}{ccccccc}
			$\Gamma_2$ & $\Psi_1$ & $\Psi_2$ & $\Psi_3$ & $\Psi_4$ & $\Psi_5$ & $\Psi_6$ \\
			\hline
			Co-1 & $(1, 0, 0)$ & $(0, 1, 0)$ & $(0, 0, 1)$ & $(1, 0, 0)$ & $(0, -1, 0)$ & $(0, 0, 1)$ \\
			Co-2 & $(1, 0, 0)$ & $(0, -1, 0)$ & $(0, 0, -1)$ & $(1, 0, 0)$ & $(0, 1, 0)$ & $(0, 0, -1)$ \\
			Co-3 & $(1, 0, 0)$ & $(0, 1, 0)$ & $(0, 0, 1)$ & $(-1, 0, 0)$ & $(0, 1, 0)$ & $(0, 0, -1)$ \\
			Co-4 & $(1, 0, 0)$ & $(0, -1, 0)$ & $(0, 0, -1)$ & $(-1, 0, 0)$ & $(0, -1, 0)$ & $(0, 0, 1)$ \\ 
		\end{tabular}
	\end{ruledtabular}
	\caption{\label{irrepD} Components of the magnetic basis vectors (in $\mu_B$) of the $\Gamma_2$ irreducible representation associated with the $(0, 1/2, 1/2)$ propagation vector at $4.25$~T (Phase D). The labeling of the four Co  the convention in \cite{Facheris2022}.}
\end{table*}

\begin{figure}
	\includegraphics[width=\columnwidth]{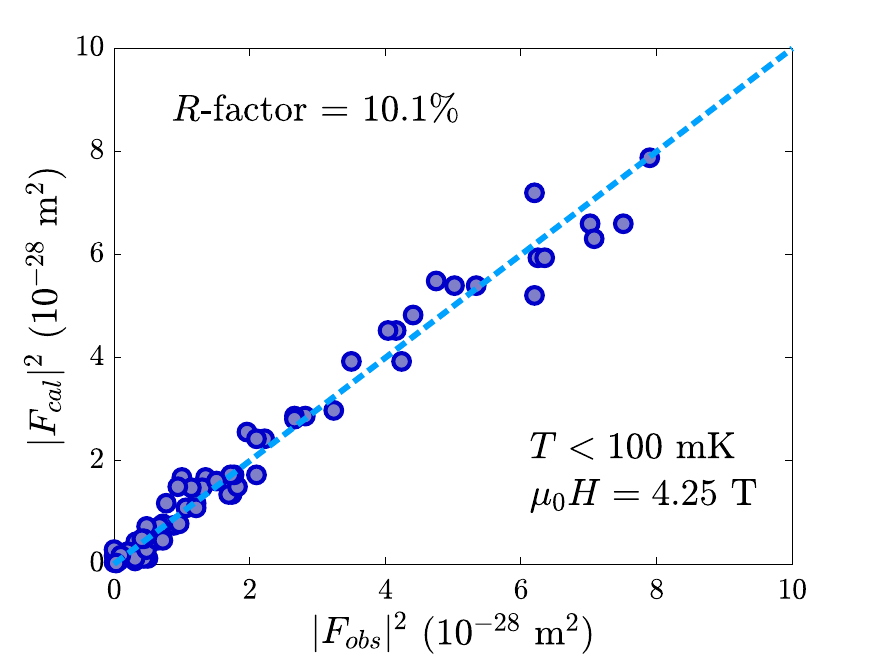}
	\caption{\label{refD} Calculated versus observed integrated intensities (from the D23 experiment) for magnetic Bragg peaks at $4.25$~T for the spin-flop structure.}
\end{figure}

A refinement of this model was performed using the FullProf Suite~\cite{FullProf_paper}.  A good fit to the data ($R$-factor $= 10.1\%$, Fig.~\ref{refD}) can be obtained with just 3 parameters $C_1 = 1.11(1)$, $C_2 = 0.01(3)$, and $C_3 = 1.14(1)$, representing the mixing coefficients of the $\Psi_1$, $\Psi_2$, and $\Psi_3$ basis vectors in the linear expansion of $\mb{A}^{(0,1/2,1/2)}_{d}$, respectively.
We have considered that $\Gamma_2$-structures can occur in several domains. Specifically, the generators of the little group produce an equivalent domain where the spins on Co-3 and Co-4 are flipped. In our analysis we included domain population as a fitting parameter. In the end, the population of the second domain was found to be negligible, which is likely due to a slight misalignment of the field away from the $\mb{b}$ axis.

 At 4.25~T the refined $(0,1/2,1/2)$-component of magnetic moment amplitude for Co-1 ions is 
	\be
	\mb{A}^{(0,1/2,1/2)}_{1} = (\, 1.11(1), 0.01(3), 1.14(1)\, )\mu_B\nonumber
	\ee 
in Cartesian coordinates. This modulation is entirely transverse to the applied field and to $\mb{b}$. Since  $\mb{b}$ is also the local anisotropy axis for each ion, we can safely assume that the uniform $(0,0,0)$-contribution to the magnetic moment is the same on all four ions, and is also parallel to the $\mb{b}$-axis. From magnetization measurements the latter is expected to be  $(0,1.94,0)~\mub$ at 4.25~T \cite{PovarovFacheris2020}. The resulting spin arrangement is the sum of the two contributions and nothing else but a spin-flop version of the zero-field structure, as illustrated in Fig.~\ref{fig:structure}(d). The spins are confined to planes that form an angle of 44$^\circ$ with the $\mb{a}$ axis. In other words, the spins lie very close to the respective easy anisotropy  planes, as one would expect. The net ordered magnetic moment is $m=2.51(2)~\mub$ on each site. 

\subsubsection{Fan phase}

The highest-field ordered Phase E with propagation vector $(0,0.47,0)$ was solved from the analysis of $25$ magnetic reflections measured at $4.8$~T in a similar manner as for Phase D. In this case, the form of the propagation vector is the same as in the SDW and UUD states \cite{Facheris2022}. The magnetic structure must belong to one of four 3-dimensional irreducible representations. Again, only one is compatible with the measured distribution of intensities, namely $\Gamma_3$, same  as that realized in the SDW and UUD states. The corresponding basis vectors are tabulated in Table~\ref{irrepE}.

\begin{table*}
	\begin{ruledtabular}
		\begin{tabular}{ccccccc}
			$\Gamma_3$ & $\Psi_1$ & $\Psi_2$ & $\Psi_3$ \\
			\hline
			Co-1 & $(1, 0, 0)$ & $(0, 1, 0)$ & $(0, 0, 1)$ \\
			Co-2 & $(1, 0, 0)$ & $(0, 1, 0)$ & $(0, 0, -1)$ \\
			Co-3 & $(e^{i\varphi}, 0, 0)$ & $(0, -e^{i\varphi}, 0)$ & $(0, 0, e^{i\varphi})$ \\
			Co-43 & $(e^{i\varphi}, 0, 0)$ & $(0, -e^{i\varphi}, 0)$ & $(0, 0, -e^{i\varphi})$ \\ 
		\end{tabular}
	\end{ruledtabular}
	\caption{\label{irrepE} Components of the magnetic basis vectors (in $\mu_B$) of the $\Gamma_3$ irreducible representation associated with the $(0, 0.47, 0)$ propagation vector at $4.8$~T (Phase E).}
\end{table*}

\begin{figure}
	\includegraphics[width=\columnwidth]{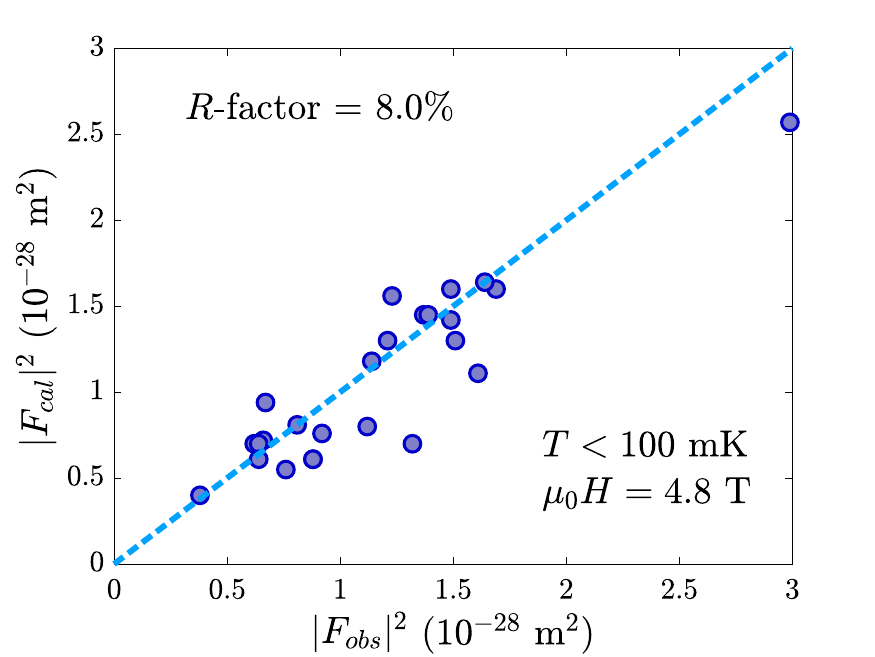}
	\caption{\label{refE} Calculated versus observed integrated intensities (from the D23 experiment) for magnetic Bragg peaks at $4.8$~T for the fan structure.}
\end{figure}

The phase shift  $\varphi=84.6^{\circ}$ in Table \ref{irrepE} is not a free parameter, but is determined by the propagation vector.  No domains are possible. This leaves three parameters needed to describe the structure, namely the mixing coefficients $C_1$, $C_2$, and $C_3$ for $\Psi_1$, $\Psi_2$, and $\Psi_3$ basis vectors from Table \ref{irrepE}, respectively. The refinement converged to a solution with $R$-factor $= 8.0\%$ (Figure \ref{refE}), yielding $C_1 = 1.18(3)$, $C_2 = 0.42(16)$ and  $C_3 = 1.00(4)$.  The amplitude of the modulated component of the Co-1 moment is 
\be
\mb{A}^{(0,0.47,0)}_{1}= (\, 1.18(3), 0.42(16), 1.00(4) \,)\mu_B.\nonumber
\ee
This oscillating component lies almost exactly in the local anisotropy plane, but is not strictly perpendicular to the applied field. Instead it forms an angle of 75$^\circ$ with it. Under the circumstances we can not make the assumption that the $(0,0,0)$ contribution is parallel to $\mb{b}$. Instead, we will assume that the length of the net magnetic moment is the same on all Co-ions, and that the  all magnetic moments are confined to the local anisotropy planes. From magnetization data, we also know the uniform part $y$-axis moment to be $2.46~\mub$ per ion. The matching spin arrangement is a tilted {\em fan structure} in which the ordered magnetic moment $m=3.26(3)~\mub$  undergoes a rotational oscillation around a fixed direction $\mb{n}$, canted by an angle of 28$^\circ$ with respect to the applied field. The angular amplitude of the oscillations is 35$^\circ$. In neighboring chains within each $J_4-J_1$ ladder the canting direction is opposite, as illustrated in Fig.~\ref{fig:structure}(e). This canting may be caused by interactions with adjacent ladders, where the spin oscillation planes are almost orthogonal. An alternative explanation would be Dzyaloshinskii-Moriya interactions.

\subsection{Neutron spectroscopy}
As mentioned in the introduction, the cryomagnet used in the experiment severely limits the access to momentum transfers in the field direction. In our cases, only reciprocal space vectors $\mb{q}=h\mb{a}^\ast+k\mb{b}^\ast+l\mb{c}^\ast$ with $-0.3<k<0.3$ can be reached. At the same time, unlike in previous studies, a wide range of momentum transfers in the $(h,0,l)$ plane are readily accessible.

\subsubsection{Paramagnetic (pseudo-polarized) phase}\label{sec:para}
\begin{figure*}
	\includegraphics[width=\textwidth]{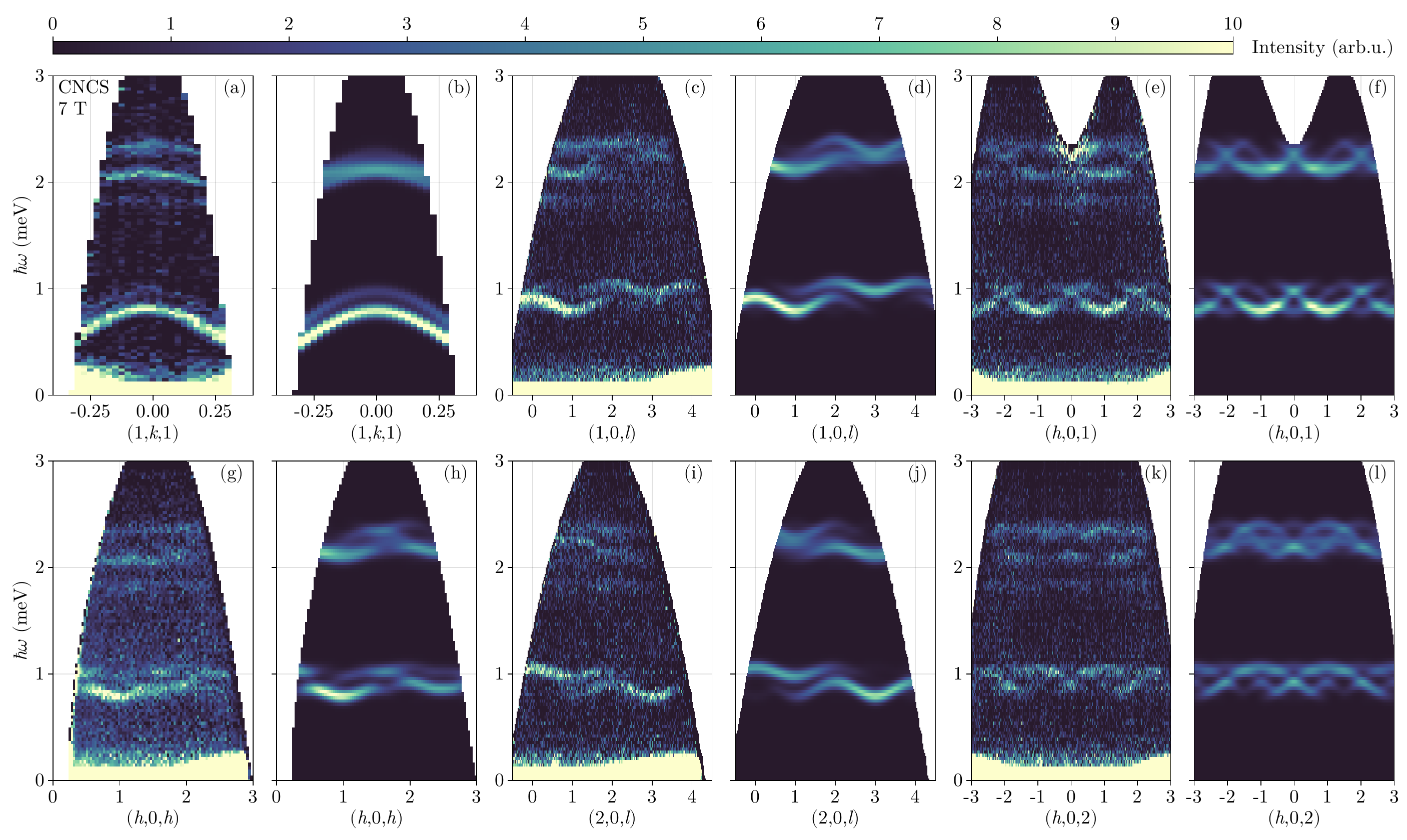}
	\caption{(a,c,e,g,i,k) Inelastic neutron scattering intensity measured in \CsCoBr in a magnetic field $\mu_0H=7$~T applied along the crystallographic $\mb{b}$ axis at $T\sim 100$~mK, in the paramagnetic pseudo-polarized phase. The data are integrated along the transverse directions in reciprocal space, as described in the text. (b,d,f,h,j,l) $SU(4)$ spin wave theory calculation based on the minimal model parameters listed in Table~\ref{table:params}.}\label{fig:paraspectra}
\end{figure*}

The data collected at the CNCS instrument in paramagnetic state at $T\sim 100$~mK and  a magnetic field $\mu_0H=7$~T applied along the $\mb{b}$ direction are shown in Fig.~\ref{fig:paraspectra}(a,c,e,g,i,k). 
Here the material is in the paramagnetic or ``pseudospin-saturated'' state \cite{PovarovFacheris2020}. The false color plots represent  measured scattered neutron intensity versus energy and momentum transfers along different reciprocal-space directions. Integration along the two transverse directions has been performed as follows: in the range $\pm 0.25$~r.l.u. for the $\mb{a}^\ast$ and/or $\mb{c}^\ast$ directions and, where applicable, $\pm 0.1$~r.l.u. along the  $\mb{b}^\ast$ axis.
These are raw data, without any background subtraction. The strong scattering at low energies is due to incoherent scattering in the sample and spurious scattering from the cryomagnet. The incident neutron energy is  $E_\text{i}=3.32$~meV.

One readily discerns two excitation sectors. At low energies we see  transitions within the lower-energy doublet of Co$^{2+}$, which is dominated by $S_z=\pm1/2$ states due to the prevailing XY-type anisotropy.
These will henceforth be loosely referred to as ``magnons''. At higher energies we see transitions between the lower and higher doublets, the latter mainly composed of $S_z=\pm3/2$ Co$^{2+}$ states. These we will denote as ``crystal field'' (CF) excitations.

\begin{table}
		\begin{tabular}{c|c|c|c|cc}
			\hline
			\hline
			 &  distance,\AA & anisotropy & value, meV & zero field & renormalized\\
			\hline
			$\bm{J_1}$ &   6.50 & $\|$ & \textbf{0.06(1)} & $BJ_1$& $B=0.55(2)$\\
			$J_2$ &   6.79 & $\bot$ &  -0.003(2) & $BJ_2$ & \\
			$J_3$ &   6.86 & $\|$ & 0.00(1)  & $B J_3$ &\\
			$\bm{J_4}$ &   7.59 & $\|$ & \textbf{0.135(5)} & $AJ_4$& $A=1.15(2)$\\
			$J_5$ &   7.65 & $\bot$ & 0.020(5)& $BJ_5$\\
			$\bm{D}$ &   -- & -- &\textbf{0.70(5)}&$D$&\\
			$\bm{d}$  &  -- & -- &\textbf{0.20(5)}&$Cd$& $C=0.40(5)$\\
			\hline
			\hline
		\end{tabular}
	\caption{Parameters of a minimal model Heisenberg Hamiltonian for \CsCoBr. The 3rd column indicates whether the corresponding bond connects magnetic ions with parallel or almost perpendicular anisotropy plane. The last column shows the renormalization of exchange constants in zero applied field.}
\label{table:params}
\end{table}

The first thing to note is that several branches show a sizable dispersion along the $\mb{c}^\ast$ axis, whereas it was very narrow in zero field. This is to be expected, since polarizing the spins in the chains by the external field removes the frustration for the zigzag inter-chain coupling. What could not be expected from  our previous weakly-coupled triangular planes model for \CsCoBr is the sizable dispersion along the $\mb{a}^\ast$ direction. It immediately shows that inter-plane interactions are at least as strong as the in-plane ones.

As demonstrated in Ref.~\cite{Coldea2002}, inelastic data collected in the saturated phase can be used to determine the true exchange constants. This is done by analyzing the measured spectra with SWT, which becomes exact in the polarized state. Strictly speaking, the latter is only true at full saturation in the axially symmetric case. In our experiment the field is applied inside the easy plane, so axial symmetry is lacking. Related to this is that a field of $7$~T only corresponds to saturation in the pseudospin model. The actual magnetization of \CsCoBr continues to increase \cite{PovarovFacheris2020}. Nevertheless, we can hope that SWT is reasonably accurate, and certainly more reliable than in zero field.

A quantitative analysis of the data was based on generalized $SU(4)$ SWT computations of the excitation spectrum. These were performed using the SUNNY software package \cite{SUNNY}. In many cases, a fit of such simulations to neutron scattering data  can be performed efficiently even in a large parameter space, as was done, for instance in Ref.~\cite{Scheie2022}. In the present case, a similar approach would be difficult to implement. The data are quite noisy. The {\em a priori} unknown and possibly structured background is often of the same order of magnitude as the weak signal. The energy resolution is barely sufficient to discern some individual excitation branches. As just mentioned, there is  no guarantee that SWT can adequately describe the system at all. Under the circumstances is difficult to even introduce a formal quantitative measure of ``goodness of fit''.
For these reasons our analysis is not a fit. Instead we used trial and error to select an {\em ad hoc} set of parameters in the proposed minimal model to reproduce the main features of the data. As itemized in the next paragraph, these include  dispersion periodicities, bandwidths and gaps (to within experimental energy resolution), as well as the qualitatively obvious intensity modulations. To enable a direct comparison with the experiment, the computation assumed a Gaussian energy resolution  of $\sigma=0.04$~meV standard deviation to roughly match that of the instrument. A reasonable overall agreement is achieved with parameter values listed in Table~\ref{table:params}. The corresponding simulated spectra are visualized in Fig.~\ref{fig:paraspectra}(b,d,f,h,j,l). The error bars are estimated conservatively: changing any parameter by the corresponding  indicated amount results in an visibly worse reproduction of the measurement.

This said, all parameters of the minimal model are well decoupled and can thus be reliably determined: 
i) The splitting between the CF and magnon sectors is primarily given by $D$. The orientation of the anisotropy plane (angle $\beta$) has only a minor effect on the spectrum in the paramagnetic state, and thus was fixed at the value quoted above. 
ii) The curvature of magnon dispersion along $\mb{b}^\ast$ is strongly influenced by $d$ and $J_4$. At the same time, the overall energy shift of the  spectrum is sensitive, in a mean field manner, to the sum of all exchange constants, but hardly at all to $d$. 
 iii) $J_2$ and $J_5$ are the only exchange pathway that provides connectivity between individual chains along the $\mb{c}$ axis, and therefore define the magnon bandwidth along $l$. Indeed,  without either $J_5$ or $J_2$ the system breaks up into disconnected triangular layers parallel to the $(b,c)$ plane. Experimentally, the magnon dispersion along $l$ is split in two branches (Figs.~6d,i) Since $J_5$ frustrates the intra-chain AF interactions and $J_2$ does not, their roles are very different. $J_5$ sets the overall bandwidths of these two modes, while $J_2$ controls their ratio. If $J_2=0$, the two bandwidths are the same. Experimentally, the upper mode has just a slightly smaller bandwidth than the lower one, which corresponds to $J_2<0$ and $|J_2|\ll J_5$. 
 iv) The magnitude of the splitting between the two magnon bands in Figs.~6d,i is set ba $J_1$ and/or $J_3$. These parameters also control the dispersion along the $\mb{a}$ direction in both the CF and magnon bands. The characteristic oscillation of intensity between the two bands along $l$  is due to the structure factor of the zigzag ladders formed by $J_4$ (legs) and either $J_1$ or $J_3$ (rungs). If these exchange constants were comparable, a contiguous triangular plane would be completed in the $(a,b)$ plane, and the intensity oscillation would disappear. This implies that one of the two exchange constants must be small. Since the two bonds are topologically equivalent, it is not possible to say which one it is, based on the excitation energies alone. As they are differently oriented, they can be told apart from the intensity modulation they produce. Unfortunately, the projections of those bonds onto the scattering plane of the experiment are quite close in length, so the effect is very subtle. Assuming $|J_1|\gg|J_3|$ gives slightly better fits, but this result can not be considered conclusive. Below we postulate that $J_1$ is dominant, which is plausible given it corresponds to nearest-neighbor Co-Co link. 
v) The final relevant parameter is the effective gyromagnetic ratio $g=2.30(5)$. It was deduced independently from the field dependence of the zone-center magnon energy measured with ESR (``p-mode'' in Ref. \cite{Soldatov2023}) and in our THz absorption experiments described below.

All in all, the agreement between the SWT calculation and experiment is rather good. In fact, the ``magnon'' part of the spectrum is reproduced almost perfectly. For CF excitations the agreement is less satisfactory. In particular, the data seem to show a weak flat band at about 1.8~meV energy transfer that cannot be reproduced by SWT with any reasonable combination of parameters.

\subsubsection{UUD phase}
\begin{figure}
	\includegraphics[width=\columnwidth]{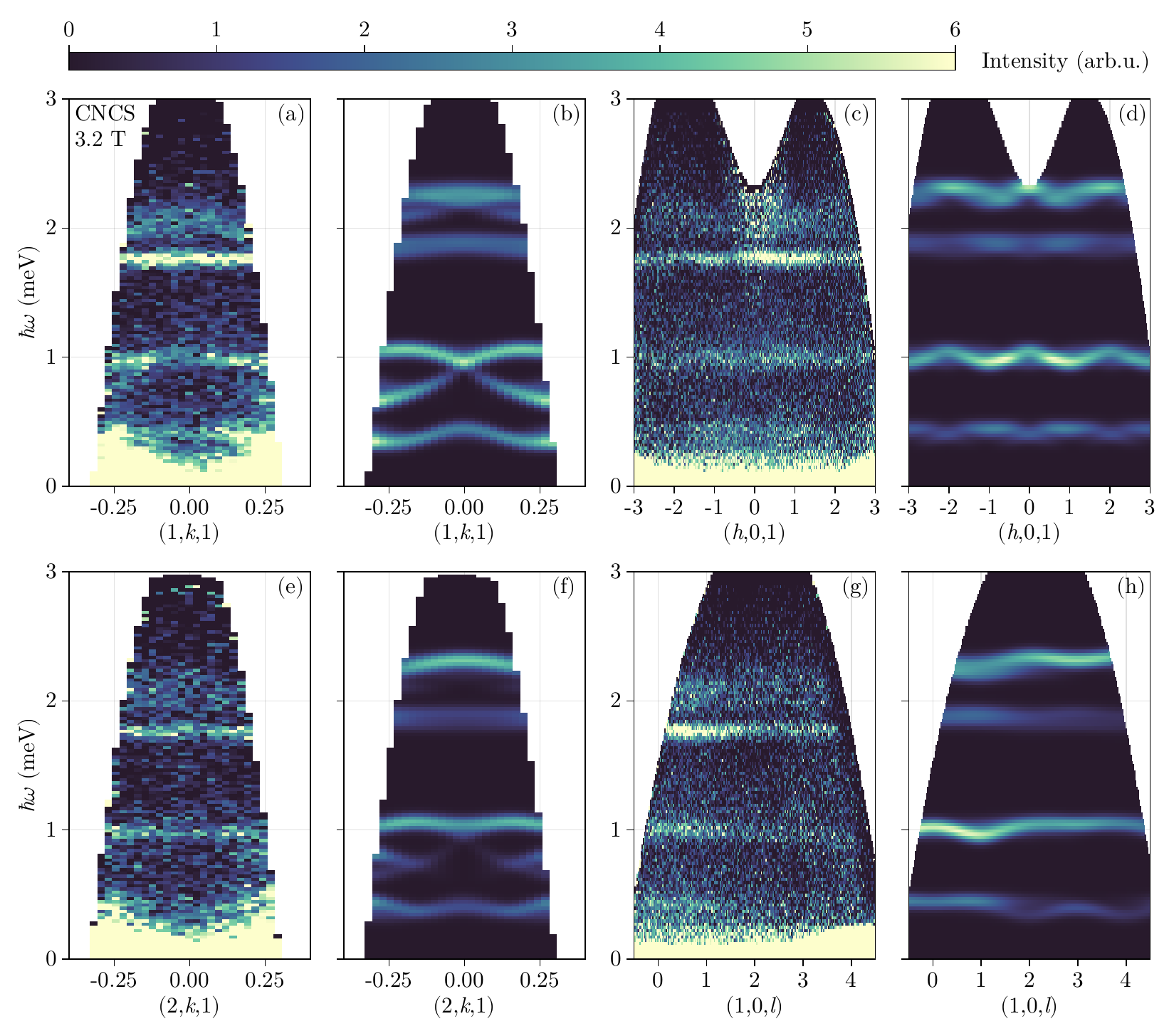}
	\caption{(a,c,e,g) Inelastic neutron scattering intensity measured in \CsCoBr in a magnetic field $\mu_0H=3.2$~T applied along the crystallographic $\mb{b}$ axis at $T\sim 100$~mK in the UUD-plateau phase. The data are integrated along the transverse directions in reciprocal space, as described in the text. The intensity units are consistent with those in Fig.~\ref{fig:paraspectra}.  (b,d,f,h) $SU(4)$ spin wave theory calculation based on the minimal model parameters listed in Table~\ref{table:params}.}\label{fig:platspectra}
\end{figure}

Inelastic neutron data collected at the CNCS instrument in the plateau Phase C at $T\sim 100$~mK and  a magnetic field of $3.2$~T applied along the $\mb{b}$ direction, that is, near the middle of the UUD plateau phase, are shown in Fig.~\ref{fig:platspectra}(a,c,e,g). The integration is done as described in Sec.~\ref{sec:para}.
The scattering intensity in the accessible wave vector transfer range is much smaller than in the saturated state shown in Fig.~\ref{fig:paraspectra} (note the different color range). Thus the data are much noisier. Still, it is very apparent that dispersion transverse to the chains practically disappears. Even the chain-axis dispersion, as seen from the narrow experimental ``keyhole'' appears almost flat.

An SWT analysis of the data is only possible if the actual magnetic structure is the ground state of the classical model. For the plateau state at 3.2~T and the proposed minimal model (Table~\ref{table:params}),  this happens to be the case. With our parameters, the plateau remains stable to 4.1~T, at which point it is replaced by the spin flop state. This is in remarkably good agreement with experiment, where the corresponding phase transition occurs at $H_\text{CD}\sim 4.0~T$ \cite{PovarovFacheris2020}. SWT calculations in the plateau phase are therefore meaningful, at least at the qualitative level. Next to each measured spectrum in Fig.~\ref{fig:platspectra} is the corresponding SWT simulation based on the high-field parameter set. Considering that this is {\em not a fit}, the agreement is not bad. The main discrepancy is the intensity distribution between the two CF branches, the lower mode being always the stronger one experimentally.

\subsubsection{Zero field}
\begin{figure}
	\includegraphics[width=\columnwidth]{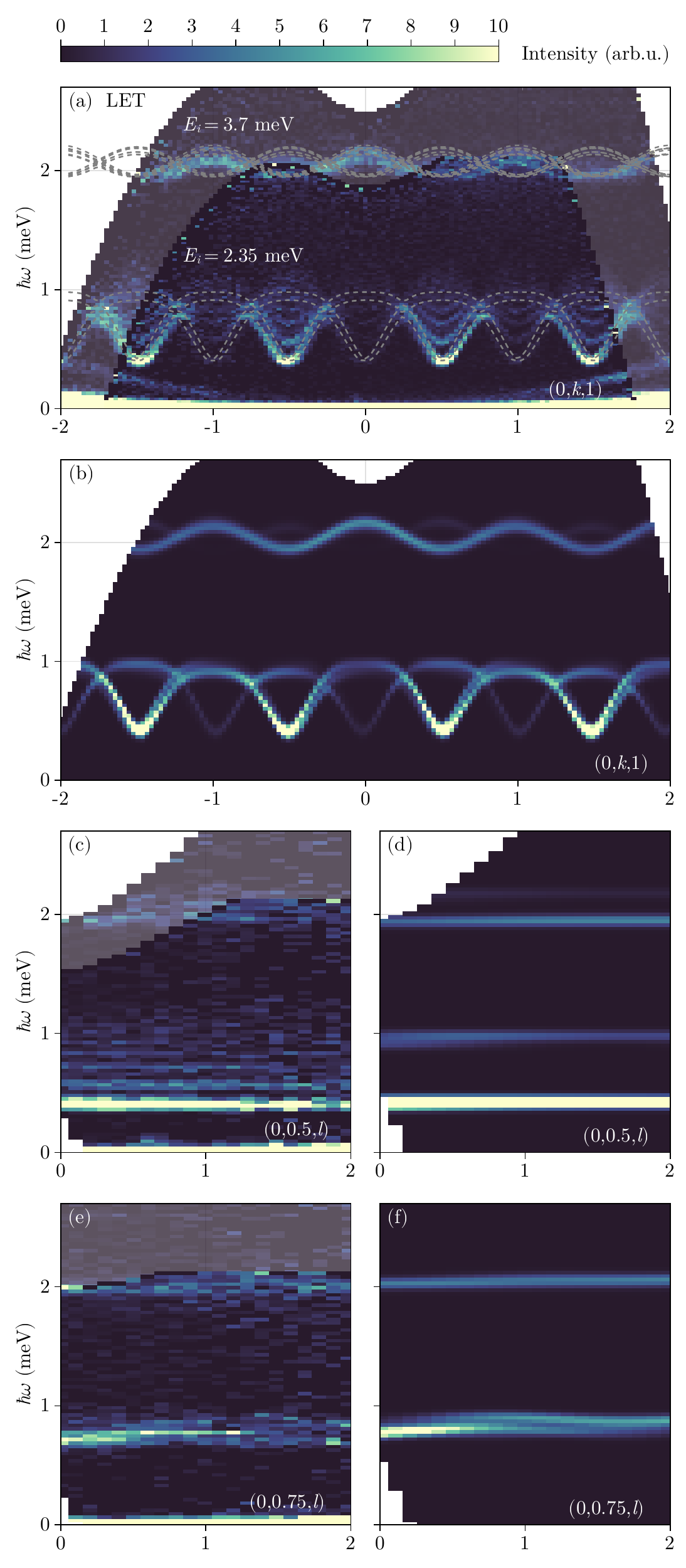}
	\caption{(a,c,e) Inelastic neutron scattering intensity measured in \CsCoBr in zero applied field at $T< 100$~mK in the Néel (stripe) phase. High- (high-contrast color map) and low-resolution (shaded) data are combined. The data are integrated along the transverse directions in reciprocal space, as described in Ref.~\cite{Facheris2023}. The extra inelastic intensity seen on the left and rights sides of (a) below 0.5~meV is spurious. It originates from the neutron beam being scattered by the cryomagnet coils.
		(b,d,f) $SU(4)$ spin wave theory calculation based on {\em renormalized} minimal model parameters listed in the last colum in Table~\ref{table:params}. Dashed lines in (a) are computed dispersion relations of all SWT modes.}\label{fig:zerospectrum}
\end{figure}

The zero field excitation spectrum (Phase A) has been previously measured in \CsCoBr on the LET spectrometer \cite{Facheris2022}. The horizontal positioning of the $\mb{b}$ axis allowed the exploration of a wide range of momenta along the chains, and provided data with a much better signal-to-noise ratio. In Fig.~\ref{fig:zerospectrum}(a,c,e) we show the same data collected with an incident neutron energy of $E_i=2.35$~meV. No background subtraction has been performed. At higher energies the same plot includes new data measured simultaneously in the same experiment with $E_i=3.71$~meV, by virtue of frame rate multiplication. This representation makes the chain-axis dispersion apparent also for the CF modes.

As already mentioned, the lower-energy part of the spectrum looks nothing like anything in SWT. This said, the lowest-energy bound state of the Zeeman ladder is essentially a single spin flip, and can be at least qualitatively associated with an SWT excitation. The CF modes can also be expected to be SWT-like. Yet SWT calculations with parameters quoted in Table~\ref{table:params} are totally incompatible with the measurements. Moreover, the Néel ``stripe'' phase with propagation vector $(0,1/2,1/2)$ that is found in \CsCoBr in zero applied fields appears not to be the classical ground state. Instead, the strong frustration in the zigzag ladders favors a helimagnetic spin arrangement, even  despite the slight easy-axis anisotropy. This is not necessarily a  contradiction. Due to the effective low dimensionality the SWT exchange parameters are expected to be strongly renormalized, as they are, for instance, in \CsCuCl \cite{Coldea2002}. For weakly-coupled spin chains the upward renormalization of the chain-axis exchange constant $J$ can be as large as $\pi/2$ \cite{Mourigal2013}. For inter-chain coupling $J_\bot$ it is downward, by a factor of the order $\sqrt{J_\bot/J}$. Indeed, the inter-chain dispersion bandwidth is  $\sqrt{JJ_\bot}$ in SWT but  of the order of $J_\bot$ in the real quantum spin model \cite{Tsukada1999}. With this in mind, we attempted to reproduce the main features of the experimental data using an SWT calculation with an effective set of parameters: $\tilde{J}_4=A\,J_4$, $\tilde{J}_{1-3,5}=B\,J_{1-3,5}$ and $\tilde{d}=C\,d$. A reasonable agreement (Fig.~\ref{fig:zerospectrum}(b,d,f)) is obtained with renormalization factors  as listed in Table~\ref{table:params}, assuming the ground state to be the stripe phase.  The dispersion of all computed modes is shown in dashed lines superimposed over the data in Fig.~\ref{fig:zerospectrum}(a).  The renormalized frustration turns out to be just small enough, yet the easy-axis anosotropy still strong enough, to make the colinear structure the classical ground state. This is inferred from the computed dispersion minima being at commensurate positions.

\subsection{THz absorption}
\begin{figure}
	\begin{centering}
	\includegraphics[width=0.95\columnwidth]{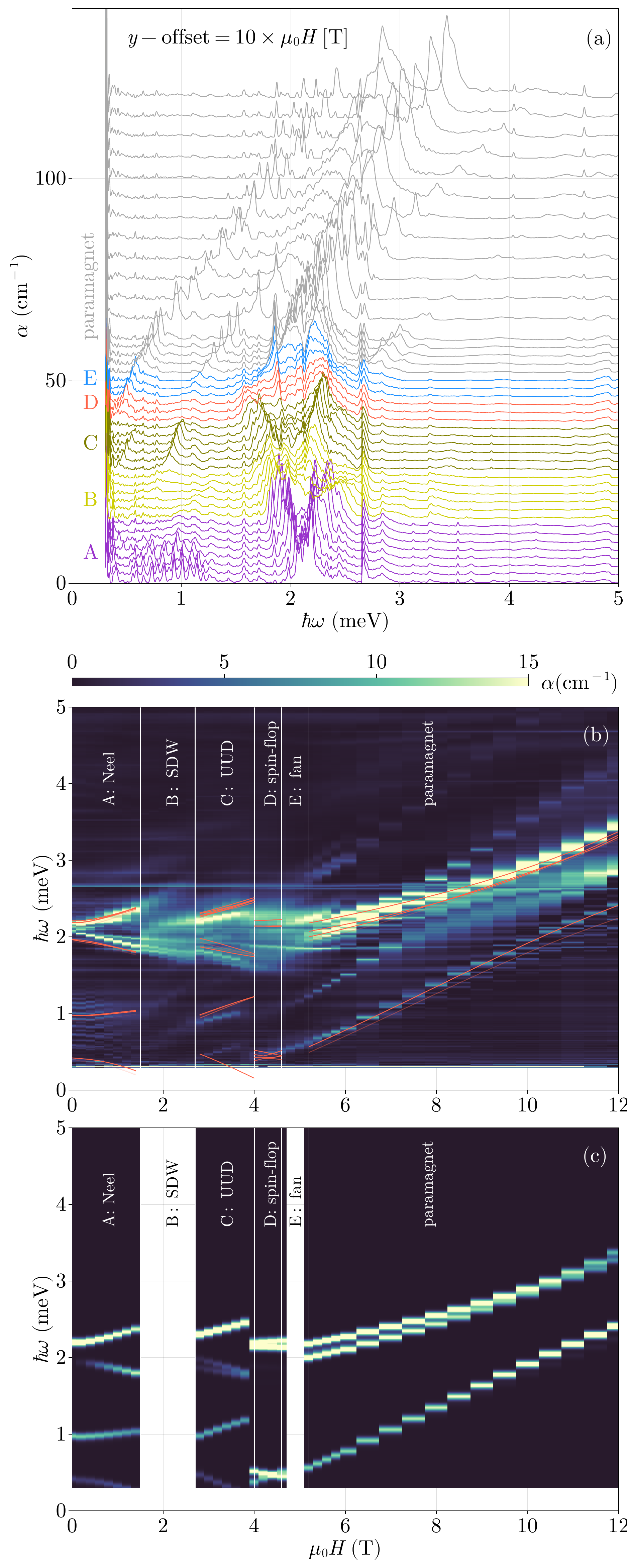}
	\end{centering}
	\caption{(a) THz absorption spectra measured at $T=200$~mK for different vales of magnetic field applied along the $\mb{b}$ axis. The various magnetic phases of \CsCoBr are labeled and color-coded. (b) Same data as a false color plot. (c) $SU(4)$ spin wave theory calculation based on minimal model parameters listed  in Tab.~\ref{table:params} for the high-field plateau and ``spin flop'' phases, and renormalized value for the low-field phase. Solid lines in (b) are energies of all SWT modes with line opacity representing the corresponding intensities. }\label{fig:thz}
\end{figure}

Previously, THz absorption experiment provided an important independent confirmation of the Zeeman-ladder type spectrum in \CsCoBr \cite{Facheris2023}. We have extended this experiment to cover a wide range of magnetic fields applied along the $\mb{b}$ axis. The corresponding absorption coefficient $\alpha$  measured at $T=200$~mK is shown in the plots in Fig.~\ref{fig:thz}(a) and (b) as a function of field and photon energy. In the false color plot, white lines indicate boundaries of field-induced magnetic phases, as determined in previous thermodynamic measurements \cite{PovarovFacheris2020}. Our results are generally in agreement with previous ESR experiments that investigated the lower-energy part of the excitation spectrum (up to about 1~meV) at a higher temperature of 0.5~K  \cite{Soldatov2023}. 

In Fig.~\ref{fig:thz}(b) the Zeeman ladder is the weak signal in the lower left part of the plot. More prominent  are sharp features in the low-energy part of the spectrum  in all other phases excluding the longitudinally modulated incommensurate SDW. In contrast, the CF excitation sector appears quite broad except in the Néel and paramagnetic/saturated phases. Another observation is the large number of distinct modes in the paramagnetic phase. 
That is rather surprising, since here we expect to be approaching the trivial regime of single-ion transitions. 
At $T>T_N$, ESR also shows complicated behavior beyond single-ion picture \cite{Soldatov2023}.

We also note some curious weakly field-dependent sharp modes that are seen at about 1.95~meV and 2.7~meV in the high field phase, but are present at all fields. These may be optical phonons.
 The ``1.95~meV'' mode shows a systematic softening with increasing field in the magnetically ordered phases.  The  ``2.7~meV'' feature is less field-dependent, but splits in two components between about 4~T and saturation, with two very narrow lines clearly discerible at 6~T. These appear to eventually anti-cross with one of the magnetic excitations. Substantial shifts of phonon frequencies in response to changing magnetic correlations in quantum magnets is a common phenomenon, with the ``pantograph'' mode in \SCB being a prominent example \cite{Bettler2020}.

Quite revealing is a comparison of the measured spectra with SWT calculations based on high-field parameters for the plateau, spin-flop and paramagnetic phases and renormalized values for the low-field stripe phase. No calculation could be performed for either incommensurate phases. The results are shown in Fig.~\ref{fig:thz}(c). The computed energies of all branches are also shown in red lines overlaying the data in  Fig.~\ref{fig:thz}(b). Line opacity is linked to the calculated absorption intensity of the corresponding mode.  The first thing we note is that several modes in the high field phase are very well reproduced. In particular, the correctly predicted slope of the lowest-energy excitations validate our choice of the $g$-factor value. At the same time, several branches are entirely missing from the calculation. No choice of parameters in the minimal model can account for these extra features in the data. In the UUD and spin flop phases the agreement between the SWT calculation and experiment is qualitative at best. In particular, the experimentally observed broadening in the CF part of the spectrum is entirely unaccounted for. In the low-field phase the behavior of the CF excitations is reproduced reasonably well, whereas a total lack of agreement for the Zeeman-ladder part is to be expected.

\subsection{Dilatometry}
\begin{figure}
	\includegraphics[width=\columnwidth]{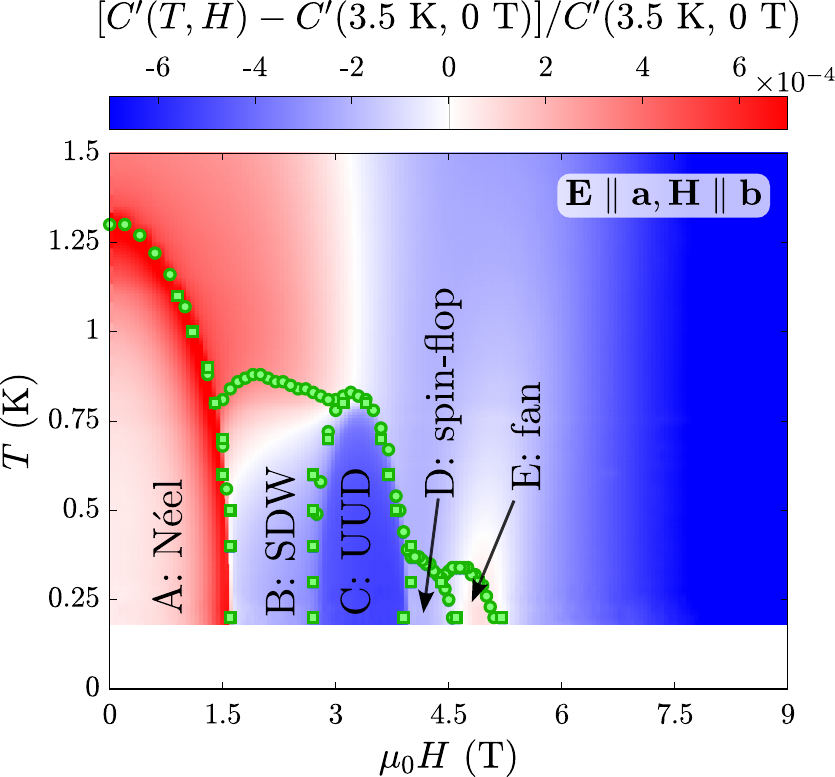}
	\caption{\label{fig:dilation} False color plot of measured change in relative capacitance $\Delta C/C$  in \CsCoBr for  $\mathbf{E}\parallel\mathbf{a}$ and $\mathbf{H}\parallel\mathbf{b}$. Green solid symbols mark magnetic phase boundaries deduced from specific heat data \cite{PovarovFacheris2020}.}
\end{figure}

While the THz spectra contain some indications of magneto-elastic coupling in \CsCoBr, dilatometric measurements make it very apparent. The measured change in sample capacitance is plotted against field and temperature in Fig.~\ref{fig:dilation}. If we assume that changes in the dielectric constant are negligible,  the observed capacitance change  is to be attributed to sample expansion and contraction. In the geometry of our experiment a positive $\Delta C$ is due to either a contraction of the sample along the $\mb{a}$ direction, or to a dilation along the $\mb{b}$ or $\mb{c}$ direction:
\be
\frac{\Delta C}{C}=\frac{\Delta b}{b}+\frac{\Delta c}{c}- \frac{\Delta a}{a}.
\ee
The entire phase diagram is clearly visible. Unfortunately, it is not possible to extract individual components of the magnetoelastic tensor from this single measurement. Further experiments will be required to fully understand the microscopic implications, but already the present data emphasize the magnitude of the magneto-elastic effect.

\section{Discussion}
The most unexpected finding of this work is the hierarchy of exchange constants. Triangular-lattice interactions that are known to dominate in isostructural Cu-based systems \cite{Coldea2002,Ono2003} play only a secondary role in \CsCoBr. Instead, the basic building blocks in our material are zigzag spin ladders, that can also be seen as zigzag chains with nearest-neighbor coupling $J_1$ and next nearest neighbor interactions $J_4$. This is a much simpler model than an extended triangular plane, particularly since anisotropy planes for all spin in each ladder are parallel. The dominant one-dimensional nature of the system also clearly favors the formation of bound states in the presence of an attractive interaction. 

The above does not, however, invalidate the conclusions of our previous work: i) At special wave vectors $\mb{qb}=\pi n$ the Zeeman ladder excitations are confined to single chains (single ladder legs in the new terminology). ii)  At all other wave vectors the bound states do disperse and therefore {\em propagate in the $c$ direction}. That must be due to the triangular lattice geometry, since other than the negligible $J_2$, it is only $J_5$ that provides any connectivity along the $\mb{c}$ axis.  iii) Other than at special wave vectors the bound states are not confined to individual chains. In view of the findings in this study, it appears most likely that these inter-chain correlations occur within individual zigzag ladders. In other words, the transverse intensity modulation of the bound states reported in Ref.~\cite{Facheris2023} may be related to the structure factor of a single ladder.

Our work once again shows that measurements in the fully saturated phase are indispensable for understanding magnetic interactions in any model compound \cite{Coldea2002}. For \CsCoBr they also highlight the importance of a complete $S=3/2$ model, rather than an effective $S=1/2$ projection. The latter corresponds to the easy plane anisotropy being infinitely strong, while in fact it is comparable in energy scale to the Zeeman energies explored in the present study. For that reason, if one were to focus on effective $S=1/2$ exchange parameters, those would have to be treated as field-dependent. Deducing true Hamiltonian parameters for the Zeeman ladder phase from high field measurements would then be entirely impossible.

The new minimal model helps make sense of the magnetic phase diagram. Even though the spin wave language is not appropriate for describing the Néel-like stripe  phase, it is still explained quasi-classically in terms of kinks. The ``spin flop'' and plateau phases are purely classical in nature appearing naturally in SWT. They are a feature of frustrated zigzag ladders with predominantly planar and additional weak easy-axis anisotropy. The main clue to the nature of the incommensurate pre-saturation phase is the field-independence of the propagation vector. This behavior has a simple classical interpretation as well. In the classical planar zigzag ladder frustration is resolved by forming a helimagnet.
In fields applied in the anisotropy easy plane this turns into an elliptical-umbrella structure or a planar spin-fan, depending on anisotropy strength. Minimizing the classical exchange energy with respect to the period of modulation for transverse spin components at a fixed longitudinal magnetization gives a propagation vector that only depends on the frustration ratio:
\be
\zeta=\frac{1}{2}-\frac{J_1}{4\pi J_4},
\ee
in the limit of $4J_4\gg J_1$. Using the numbers from Table~\ref{table:params} gives $\zeta=0.465$, almost exactly as seen experimentally. We conclude that the only ``truly quantum'' phase in \CsCoBr is the incommensurate SDW state.

It has to be re-iterated that our SWT analysis has its limitations. The agreement of minimal model simulations with experiment is not perfect, particularly in what concerns the CF sector. Part of it may be due to  quantum effects beyond the SWT paradigm and part to our simplified treatment of magnetic anisotropy, where everything is reduced to a single-ion term. The discrepancies in the high field regime are the most telling, since here the SWT is expected to work best. It is well known that extended SWT sometimes predicts excitations such as longitudinal modes that are not reproduced experimentally  (see, for example, Refs.~\cite{Mannig2018,HayashidaMatsumoto2019}) due to their decay into transverse excitations \cite{ZhitomirskyChernyshev2013}. In our case though, there seem to be more modes visible in the measurements than predicted. It can not be excluded that some essential component is still missing from our simple model for \CsCoBr. As discussed above, one such component may be the very strong magneto-elastic coupling that seems to affect not just bulk properties, but also the excitations. A 0.1\% change in lattice parameters may not appear very large, but the magnetoelastic  effect on exchange constants via  bond angles close to the Kanamori-Goodenough critical value may be considerable. Further studies in that direction are needed.

\section{Conclusion}
Particularly for frustrated systems, the spin Hamiltonian of a material can not be considered known until neutron spectroscopy has been performed in the high field paramagnetic state. \CsCoBr should be seen as a network of coupled frustrated zigzag spin ladders, rather than a distorted triangular lattice system. This opens the possibility to quantitatively study the exotic spin excitations in this compound using matrix product states numerical methods, which are known to be exceptionally effective in application to quasi-1D systems. At the same time triangular-lattice inter-ladder interactions in \CsCoBr are significant and lead to new physics compared to that of a single ladder. A new dimension to the problem is brought by strong magneto-elastic interactions that will need to be explored more thoroughly in \CsCoBr.

\acknowledgements
One of the authors (AZ) would like to thank Prof. Giamarchi (University of Geneva) and Prof. Coldea (University of Oxford) for illuminating discussions. This research was in part funded by a MINT grant ``Tricoordinated frustrated quantum magnets'' of the Swiss National Science Foundation. It made use of resources at the Spallation Neutron Source, a DOE Office of Science User Facility operated by the Oak Ridge National Laboratory. We acknowledge support by
the Estonian Research Council Grants No. PRG736 and
No. MOBJD1103, and by European Regional
Development Fund Project No. TK134. Experiments at the ISIS Neutron and Muon Source were supported by beam time allocation from the Science and Technology Facilities Council (DOI: \cite{WISH_data,LET_data}). This research benefited from neutron beam time at the instrument D23 at the Institute Laue Langevin provided via the Collaboration Agreement between the French Alternative Energies and Atomic Energy Commission (CEA) and the École Polytechnique Fédérale de Lausanne (EPFL) in Switzerland and funded by the Swiss State Secretariat for Education, Research, and Innovation (SERI).

\end{document}